\documentclass{article}

\usepackage{microtype}
\usepackage{graphicx}
\usepackage{subcaption}
\usepackage{booktabs} 

\usepackage{hyperref}

\usepackage[accepted]{icml2026}

\usepackage{amsmath}
\usepackage{amssymb}
\usepackage{mathtools}
\usepackage{array} 
\usepackage{amsthm}
\usepackage{amsfonts}
\usepackage{nicefrac}

\usepackage[capitalize,noabbrev]{cleveref}

\usepackage{caption}
\usepackage{float}
\usepackage{multirow}
\usepackage{longtable}
\usepackage{mdframed}
\usepackage{xcolor}

\definecolor{unsupportedred}{HTML}{C0392B}
\definecolor{weakorange}{HTML}{E67E22}
\definecolor{plausibleblue}{HTML}{1F77B4}
\definecolor{verifiedgreen}{HTML}{2CA02C}
\definecolor{sectionblue}{HTML}{2C3E50}
\definecolor{lightgray}{HTML}{F4F6F8}
\definecolor{midgray}{HTML}{BDC3C7}
\definecolor{commentbg}{HTML}{E8EEF5}
\definecolor{posgreenbg}{HTML}{E9F5E9}
\definecolor{negredbg}{HTML}{F9EBEB}

\newcommand{\answerTODO}[1][]{\textcolor{red}{\bf [TODO]}}

\newcommand{\verdictbadge}[2]{%
  \colorbox{#1}{\color{white}\bfseries\tiny\ #2\ }%
}

\theoremstyle{plain}

\theoremstyle{definition}

\theoremstyle{remark}

\usepackage[textsize=tiny]{todonotes}

\icmltitlerunning{Submission and Formatting Instructions for ICML 2026}

\begin{document}

\twocolumn[
  \icmltitle{\emph{DeepRoot}: A KG-Coordinated Multi-Agent System for Therapeutic Reasoning over Historical Medical Texts}



  \icmlsetsymbol{equal}{*}

  \begin{icmlauthorlist}
    \icmlauthor{Zijian (Carl) Ma}{equal,dept,sch}
    \icmlauthor{Sean J. Wang}{equal,dept,sch}
    \icmlauthor{Sijbren Kramer}{equal,dept,sch}
    \icmlauthor{Li Erran Li}{comp}

  \end{icmlauthorlist}

  \icmlaffiliation{dept}{Department of Bioengineering}
  \icmlaffiliation{sch}{Stanford University}
  \icmlaffiliation{comp}{Amazon AWS AI}

  \icmlcorrespondingauthor{Zijian (Carl) Ma}{mazijian@stanford.edu}

  \icmlkeywords{Machine Learning, ICML}

  \vskip 0.3in
]



\printAffiliationsAndNotice{\icmlEqualContribution}  

\begin{abstract}
Historical medical archives and traditional medicines hold immense potential for drug discovery and remain a primary source for current drug development. However, pre-ontological prose and idiosyncratic taxonomies prevent the standardization and medical modernization of the data for use in current biomedical pipelines. Furthermore, no existing LLM agent system, whether tool-calling, retrieval-augmented, or agentic deep-research, can convert such text into verifiable drug-discovery leads at scale. We close this gap with DeepRoot, a multi-agent LLM system that jointly builds and utilizes a verified knowledge graph, showing that grounding and reasoning---often conflated---are separable axes the system can compose for therapeutic reasoning. Applied to the \emph{Shen Nong Ben Cao Jing}, DeepRoot recovers $10$ of $21$ held-out compound--disease treatment pairs at R@$20$ ($47.6\%$ vs.\ $4.8\%$ for a raw corpus LLM and $\sim\!2.4\%$ random) and dominates an LLM-as-judge audit for reasoning quality over baseline LLMs and LLMs with direct tool-call access to the same APIs DeepRoot itself
queries. Tool-using LLMs hallucinate evidence on $87\%$ of claims, versus $7$--$10\%$ for DeepRoot. Graph-only inference hallucinates $0\%$ but ranks lowest on reasoning coherence; DeepRoot KG\,+\,LLM is the only condition to win on both axes, pointing toward a route for systematic mining and repurposing of historical medical knowledge.
\end{abstract}

\section{Introduction}

Natural products---chemical compounds synthesized by living organisms---remain the leading source of approved drugs and provide scaffolds for
developing more potent derivatives~\citep{newman2020natprod,
koehn2012biosynthetic}. Many natural products have been uncovered through
mining traditional medicines, including morphine from opium poppies and
the antimalarial artemisinin, with the latter isolated by Tu Youyou
after consulting a 4th-century Chinese medical text~\citep{tu2011artemisinin,
brook2017morphine}.

ML, DL, and LLM approaches for mining historical medical
texts at scale have been reported before but treat the text as pure input-classification problems without a reasoning trace grounded in
verified biological evidence or mechanism ontologies~\citep{li2024evibert, hui2020tcmbert, liu2025tcmllm, dai2024tcmchat}. In parallel, multi-agent LLM systems leverage a shared knowledge graph
(KG) for coordinated reasoning~\citep{ghafarollahi2024sciagents,
rasmussen2025zep}, but only qualitatively: they neither ablate the
graph against agent decomposition, nor evaluate on the regimes we
target---historical clinical cases where traditional text lacks clean
ontological anchors, and discovery problems with sparse ground truth. 

\begin{figure}[h]
    \centering
    \includegraphics[scale=0.35]{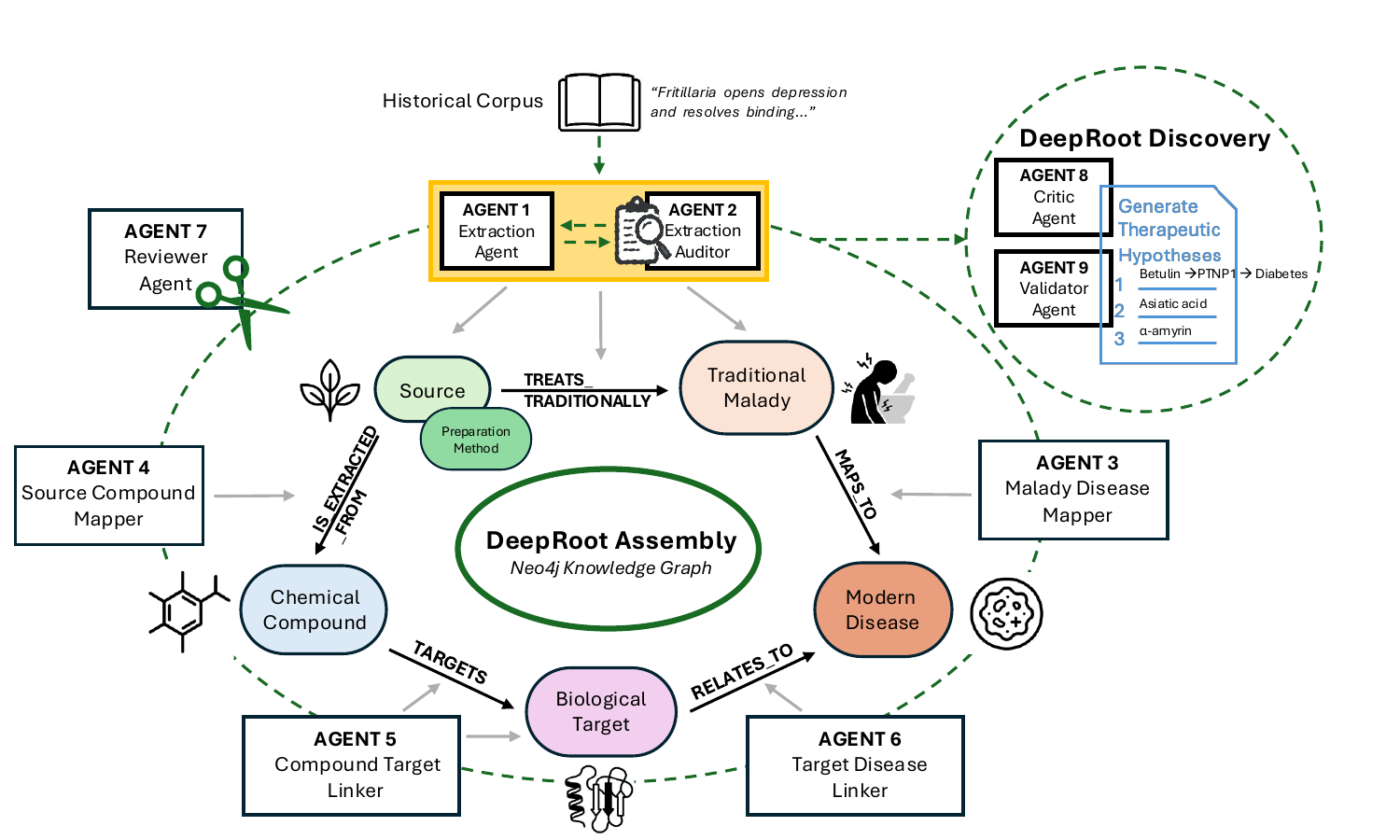}
    \caption{Schematic of DeepRoot. Graph nodes and edges are represented by rounded rectangles and black arrows. Gray arrows indicate creation of specific nodes and edges by particular agents.}
    \label{fig:overview}
\end{figure}

Building on these advances, we introduce \textbf{DeepRoot} (Figure~\ref{fig:overview}), a
multi-agent LLM pipeline where agents collectively construct and reason
over a shared KG (Neo4j). Closest to our work is OpenTCM~\citep{he2025opentcmgraphragempoweredllmbasedtraditional}, which uses a Graph-RAG architecture for LLM reasoning. However, its construction relied on expert oversight and pure LLM-generated outputs. \textbf{DeepRoot Assembly} agentically populates the knowledge
graph via seven specialized agents that combine LLM canonicalization
with strict verification against curated biomedical databases.
\textbf{DeepRoot Discovery} then employs critic and discovery agents, leveraging Neo4j Cypher walks for subgraph traversal to evaluate therapeutic claims and identify potential therapeutics that are mechanistically grounded in the KG.

\section{Methods and KG construction}

\subsection{Dataset and grounding sources}
\label{sec:dataset}

\textbf{Corpus.} We evaluate on the \emph{Shen Nong Ben Cao Jing} materia medica, segmented into
71 chunks. The corpus catalogues plants, animals, and minerals
(\emph{sources}), maladies, preparation
methods, and claimed therapeutic uses.

\textbf{External grounding.} Every entity is verified against curated
biomedical databases. Sources are linked to compounds via
COCONUT2.0~\citep{sorokina2024coconut} (natural products) and
PubChem~\citep{kim2023pubchem} (chemicals); compounds are linked to
molecular targets and clinical indications via
ChEMBL~\citep{mendez2019chembl}; protein targets are linked to diseases via
Open Targets~\citep{ochoa2021opentargets}, and pathogenic-organism targets
via NCBI Taxonomy~\citep{schoch2020ncbitax} with OLS4~\citep{ebi_ols4}. Modern
disease nodes are anchored to ICD-10, MeSH, SNOMED, MONDO, and DOID
identifiers via NLM and EBI lookup services. 

\subsection{Knowledge graph schema}
\label{sec:schema}

The graph has six node types and seven edge types
(Figure~\ref{fig:overview}). A therapeutic claim is \emph{verifiable}
when its mechanistic loop closes: a \texttt{Source} treats a
\texttt{Traditional\_Malady} that maps to a \texttt{Modern\_Disease};
the source contains a \texttt{Chemical\_Compound} that targets a
\texttt{Biological\_Target} which itself relates to that same
\texttt{Modern\_Disease}. Identity for compounds is the
RDKit-computed InChIKey and identity for targets is the curated
ChEMBL ID, so equivalent entities arriving from different routes
collapse onto the same node. Full schema is tabulated in Table ~\ref{tab:kg-schema-full}.

\textbf{Assembly.} Seven specialized agents populate the graph in
dependency order: an \emph{extractor} emits Source, Malady, and
Preparation nodes from raw text; an \emph{auditor} canonicalizes
sources and archives evidence spans that fail substring verification
against their source chunk; three \emph{linkers} ground audited
entities to compounds, molecular targets, and target-to-disease
associations using the databases above; a \emph{malady-to-disease
mapper} follows a generate-then-verify protocol in which LLMs propose
canonical names and ontology codes are recovered only by tolerant
exact match, eliminating hallucinated
identifiers; and a \emph{reviewer} archives orphans and off-domain
entities.

\textbf{Resulting graph.} On \emph{Shen Nong Ben Cao Jing}, Assembly
yields \textbf{21{,}111 active nodes} (415 sources, 294 maladies, 129
modern diseases, 18{,}012 compounds, 2{,}211 targets, 50 preparations)
and \textbf{52{,}467 active edges} (32{,}909 \texttt{IS\_EXTRACTED\_FROM},
16{,}696 \texttt{TARGETS}, 1{,}841 \texttt{RELATES\_TO}, 431
\texttt{TREATS\_TRADITIONALLY}, 257 \texttt{MAPS\_TO}, 301
\texttt{KNOWN\_TREATS}, 32 \texttt{PREPARED\_AS}). A visual example of nodes originating from a single extracted source is presented in Figure~\ref{fig:neo4j-example}.

\section{Results}

\begin{figure*}[t]
\centering
\includegraphics[scale=0.395]{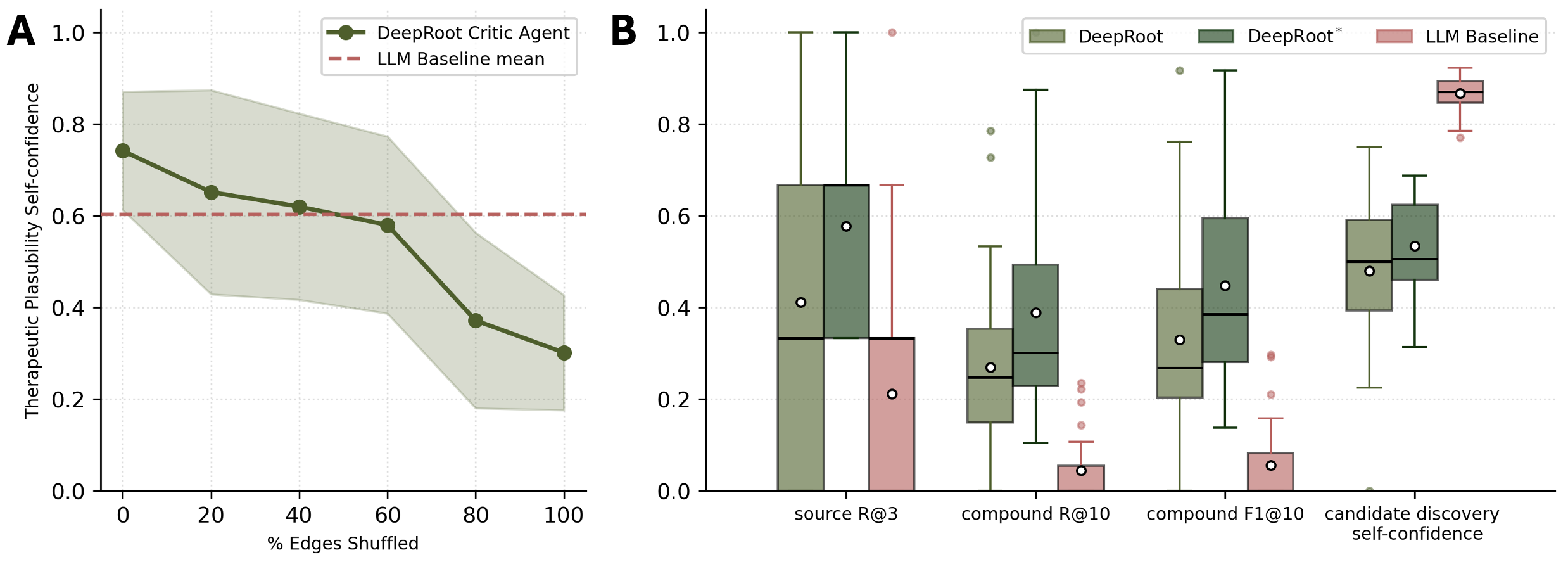}
\caption{\textbf{(A)} Critic agent self-reported confidence for the therapeutic plausibility of each source-text pair vs. KG edge-shuffle fraction (n=30 source-text pairs). 
\textbf{(B)} Source-and-compound recovery of DeepRoot Discovery, LLM baseline (both using Gemini3.1 Flash Lite). ${\text{*}}$Batch evaluation by processing all $30$ mini-corpora in a single invocation. Candidate discovery self-confidence refers to the mean self-reported confidence in each model's proposed compound candidates per mini-corpus (n=30 mini-corpora).}
\label{fig:eval}
\end{figure*}
\subsection{Knowledge-graph ablation: edge perturbation tests structural dependence}
\label{sec:eval-perturbation}

First, to verify that DeepRoot Discovery genuinely relies on graph structure, we progressively shuffled the graph edges and tasked the critic agent with evaluating 30 extracted closed-loop
source--malady claims. As expected, the Critic's self-confidence in the therapeutic plausibility of the source based on the text decreases as edge perturbation increases, demonstrating responsiveness to the KG's integrity (Figure~\ref{fig:eval}A). Around $50\%$ perturbation,
the critic's confidence converges with the raw LLM baseline, suggesting
that the KG signal has been degraded enough that the critic behaves
similarly to an LLM without structured graph support. Furthermore, past $50\%$, the score continues to decrease to ~0.30, reflecting KG-dependent scoring. 

\subsection{KG-guided recovery of mechanistically supported candidates}
\label{sec:eval-recovery}

Next, we tested whether DeepRoot can use the KG to recover
mechanistically grounded candidates from noisy historical text. For this, we synthesized evaluation cases by selecting sets of 3 closed-loop and 7 non-closed-loop distractor sources. The associated paragraphs of those sources were then interweaved into a mini-corpus and fed to different models to rank the sources and candidate chemical compounds (Figure~\ref{fig:eval}B). We report source recall@3, compound recall@10 (the fraction of closed-loop compounds recovered within the top-10 candidates), and mean self-confidence (0–1) related to the therapeutic plausibility of candidate compounds. Because each passage may contain $\sim500$ compounds, compound recall@10 directly tests whether KG-grounded scoring concentrates the likely leads. 

Over $30$ mini-corpora, DeepRoot Discovery outperforms
the LLM baseline, achieving $1.95\times$ higher source recall and
$6.11\times$ higher compound recall (Figure~\ref{fig:eval}B). Surprisingly, despite DeepRoot Discovery being theoretically capable of fully traversing the KG, recovery was not perfect. This is likely due to two factors: framing evaluation as an inference task~\citep{edwards2024lmtextclass} and incomplete subgraph traversal, which could explain why batch evaluation across all mini-corpora improved all metrics since it could indirectly surface shared relevant paths for the Critic. 

Nevertheless, the recall and F1 gains validate that KG augmentation meaningfully enhances parsing and ranking. Notably, the LLM baseline overstates therapeutic relevance, with a self-confidence of $0.87$, versus DeepRoot Discovery's $0.48$, which closely aligns with the latter's source
recall@$3$ of $0.41$ (Figure~\ref{fig:eval}A). This alignment
suggests that self-confidence in a KG-augmented system is effectively
bounded by retrieval accuracy. In contrast, the other modalities demonstrated high self-confidence hallucinations, which is a phenomenon previously reported for both LLMs and agents~\citep{lin2022truthfulqa, Lin2025LLMbasedAS}. Together with the KG ablation study, we establish that KG contributes meaningfully to the reasoning capabilities of an underlying LLM.

\subsection{Blind rediscovery of held-out validated treatments}
\label{sec:compound-recovery}

Whereas Section~\ref{sec:eval-recovery} tested whether DeepRoot
surfaces \emph{mechanistically grounded} candidates---compounds for
which the graph itself closes a compound$\to$target$\to$disease
loop---this experiment tests whether the system can blind-rediscover \emph{empirically validated}
compound--disease treatments after we hide them. Concretely, for
each held-out pair (a \texttt{KNOWN\_TREATS} edge sourced from
ChEMBL clinical indications) we delete the edge and all
stereochemical siblings (planar-InChIKey prefix) from the
validator, then ask DeepRoot Discovery to re-rank candidates for
the disease. We evaluate on a $21$-pair historical set,
and compare against a raw-corpus LLM given the full
\emph{Shen Nong Ben Cao Jing} and asked to rank the same top-$K$ (Table~\ref{tab:compound-recovery}).

DeepRoot Discovery recovers 10 of 21 held-out pairs (R@20 = 47.6\%), compared with the LLM baseline
(1 of 21, R@20 = 4.8\%). Per-disease candidate pools span $87$--$1{,}954$ compounds
(median $835$), so random $R@20 \approx 2.4\%$, suggesting that the result is far above random retrieval. 

\begin{table}[H]
  \caption{Held-out \texttt{KNOWN\_TREATS} recovery on $21$
  historically reachable ChEMBL indication pairs. R@$k$ in \%,
  MRR unitless.}
  \label{tab:compound-recovery}
  \centering
  \small
  \setlength{\tabcolsep}{4pt}
  \begin{tabular}{lccccc}
    \toprule
    Method & R@1 & R@5 & R@10 & R@20 & MRR \\
    \midrule
    DeepRoot Discovery & 9.5 & 28.6 & 33.3 & \textbf{47.6} & \textbf{0.161} \\
    Raw-corpus LLM  & 0.0 & 4.8  & 4.8  & 4.8           & 0.012 \\
    \bottomrule
  \end{tabular}
\end{table}

\begin{table*}[!t]
  \caption{Reasoning-quality evaluation: seven conditions graded by
  Claude Sonnet 4.6 over a stratified sample of $30$
  source$\to$malady claims. Scores are means on $[1,5]$; Hallu.\ is
  the rate of the judge's \texttt{hallucinated\_evidence} flag in
  $[0,1]$. \textbf{Bold} = best per column.}
  \label{tab:taskA-judge}
  \centering
  \small
  \setlength{\tabcolsep}{3pt}
  \renewcommand{\arraystretch}{0.95}

  \small{%
  \begin{tabular}{llcccccccc}
    \toprule
    System & Components
      & Overall$\uparrow$ & EF$\uparrow$ & VA$\uparrow$ & RC$\uparrow$
      & CM$\uparrow$ & UC$\uparrow$ & AC$\uparrow$ & Hallu.$\downarrow$ \\
    \midrule
    DeepRoot --- Gemini 3.1 Pro & graph + LLM
      & \textbf{3.83} & 4.53 & 4.47 & \textbf{3.97}
      & \textbf{4.07} & \textbf{3.73} & \textbf{3.67} & 0.10 \\

    DeepRoot --- Gemini 2.5 Flash & graph + LLM
      & 3.77 & \textbf{4.67} & 4.37 & 3.83
      & 3.63 & 3.57 & 3.63 & 0.07 \\

    DeepRoot --- Gemini 3.1 Flash Lite & graph + LLM
      & 3.70 & 4.60 & 4.27 & 3.73
      & 3.70 & 3.60 & \textbf{3.67} & 0.07 \\
    \midrule

    Graph-only & graph, no LLM
      & 3.55 & 4.55 & \textbf{4.55} & 2.69
      & 3.21 & 3.31 & 2.93 & \textbf{0.00} \\

    Text + LLM (G3.1 FL) & corpus + LLM
      & 3.17 & 3.10 & 2.80 & 3.47
      & 3.67 & 3.27 & 3.17 & 0.13 \\

    Tool-call + LLM (G3.1 FL)
      & {\tiny ChEMBL/OT/PubMed/MeSH}
      & 2.47 & 2.30 & 2.97 & 2.80
      & 3.30 & 2.63 & 2.70 & 0.87 \\

    \bottomrule
  \end{tabular}%
  }\\[0.3em]
  {\footnotesize
  EF: evidence fidelity; VA: verdict alignment; RC: reasoning coherence;
  CM: clinical mapping; UC: uncertainty calibration; AC: actionability;
  Hallu.: hallucination rate.}
\end{table*}

\subsection{Benchmarking DeepRoot's therapeutic reasoning against diverse baselines}
\label{sec:eval-judge}

We audit critic-agent outputs with an independent LLM judge
(Claude Sonnet 4.6, cross-family from the graded systems) on $30$
stratified source--malady claims across seven conditions
(Table~\ref{tab:taskA-judge}): the DeepRoot Discovery at three LLM
tiers (Gemini 3.1 Pro / 2.5 Flash / 3.1 Flash Lite), a
graph-only baseline (no LLM), an LLM-only baseline given corpus
passages, and a tool-call
LLM with direct access to the same APIs (ChEMBL, Open Targets,
PubMed, MeSH) that DeepRoot Assembly itself queries. The judge
scores six dimensions on $[1,5]$ and flags hallucinated evidence
per claim.

All three KG-augmented configurations outperform every baseline on
overall score. Even DeepRoot--Lite ($3.70$) exceeds both the graph-only
condition ($3.55$) and the tool-calling LLM ($2.47$). This contrast
highlights a tradeoff between grounding and synthesis. The tool-calling
agent triggers the judge's hallucinated-evidence flag on $87\%$ of claims,
despite having on-the-fly access to the same set of APIs. By contrast,
the graph-only condition produces no hallucinated evidence by construction,
but exhibits the weakest reasoning coherence ($2.69$). KG-augmented LLMs
therefore occupy a favorable middle ground: they maintain low hallucination
rates ($7$--$10\%$) while preserving the reasoning and synthesis capacity
absent from graph-only scoring. 

\subsection{Human-expert evaluation of DeepRoot reasoning as a qualitative case study}

To evaluate the reasoning quality of DeepRoot Discovery as a traditional medicine knowledge assistant, we constructed a mini-corpus of 50 randomly sampled source–malady pairs from \textit{Shen Nong Ben Cao Jing}, each paired with its associated textual evidence. This setting reflects a potential use case in which a scientist seeks to assess whether observed historical claims about a source’s therapeutic potential are grounded in modern biological evidence.

\begin{table}[H]
\caption{Classified modern disease agreement and verdict agreement with reference to DeepRoot. Verdict refers to the system-interpreted therapeutic plausibility of a source--malady pair.}
\label{tab:verdict-distribution}
\centering
\small
\setlength{\tabcolsep}{4pt}

\begin{tabular}{@{}lcccccc@{}}
\toprule
System
& DA
& VA
& PR
& PB
& W
& U \\
\midrule

DeepRoot --- (G3.1 FL)
& -- & -- & 3 & 39 & 4 & 4 \\

Biomni (w/ full agent env.)
& 42 & 30 & 0 & 32 & 12 & 6 \\

Text + LLM (G3.1 FL)
& 33 & 13 & 1 & 11 & 19 & 19 \\

\bottomrule
\end{tabular}

\vspace{0.3em}
{\footnotesize
DA: disease agreement (\% of source--malady pairs mapped to the same modern disease as DeepRoot); VA: verdict agreement (\% of pairs assigned the same verdict as DeepRoot); PR: Previously Reported; PB: Plausible or Better (Previously Reported, Very Plausible, or Plausible); W: Weak; U: Unsupported.
}
\end{table}

Comparing DeepRoot with Biomni (a biomedical reasoning agent environment) ~\citep{huang2025biomni}, we find good alignment of disease classification and verdicts (Table ~\ref{tab:verdict-distribution}). We provide full responses from DeepRoot and Biomni Appendix~\ref{app:examples}. Overall, DeepRoot Discovery clearly leverages the KG for reasoning. Biomni also shows strong reasoning and tool-calling capabilities. Notably, both DeepRoot and Biomni cite the same biological targets for their reasoning in example~1. But example~8  highlights a key limitation of KG over-reliance. Although DeepRoot's underlying LLM  identified that the hydrolyzed version of certain compounds were bioactive, those downstream products were absent from the KG, preventing KG path completion and leading to low confidence despite biological plausibility.

\section{Discussion and Conclusion}
\label{sec:conclusion}

DeepRoot shows that historical materia medica can be converted from
pre-ontological prose into an auditable biomedical knowledge graph that
supports mechanistic therapeutic reasoning. On the \emph{Shen Nong Ben
Cao Jing}, this construction pass enables held-out treatment recovery
and substantially lower hallucinated-evidence rates than LLM-only,
tool-calling, and biomedical-agent baselines; with broader historical
corpora, the same framework could support larger-scale drug repurposing,
de novo therapeutic candidate nomination, and prioritization of
experimentally testable natural-product hypotheses. 

It is also important to highlight the gap between DeepRoot
(LLM + KG) and LLMs given direct access to the same
biomedical APIs. Our results suggest that building a verified knowledge
graph suppresses hallucination in a way that querying those resources
at inference time does not. We show that building a verified knowledge graph suppresses hallucination in a way where querying those same APIs at inference time does not. For corpora that predate modern ontologies, retrieval-augmented and tool-using agents need a construction pass first, rather than on-demand calling. The same pattern may transfer
to other historical materia medica, including Ayurvedic and broader
ethnopharmacological archives, as well as structured ranking problems
beyond traditional medicine. Furthermore, DeepRoot Assembly is a one-time invocation costing  $\sim\$0.25$/corpus and avoiding the recurring expense of on-demand database retrieval. In these settings, an agentically
constructed KG offers the additional advantage that new curated claims,
user submissions, and external evidence can be incorporated over time,
expanding coverage while preserving traceability for future candidate
ranking. 

\paragraph{Limitations.}
\label{sec:limitations}
\emph{(i)~Corpus:} a single $71$-chunk materia medica; transfer to
other historical corpora is unverified. \emph{(ii)~Sample size:} the
held-out slice is $N=21$ pairs, single seed, no bootstrap CIs.
\emph{(iii)~Priors:} flat, face-validity, uncalibrated
(Appendix~\ref{app:priors}). \emph{(iv)~Coverage:} Open Targets is
human-disease-only, leaving non-modern indications unscored.
\emph{(v)~DeepRoot Discovery reasoning:} rationale quality is bounded by the
underlying LLM. \emph{(vi)~Comparasion evalutions with other LLM modalities:} While we reported the best result for LLM or biomni from prompt engineering, limited analysis was placed in this area, but previous studies have demonstrated that such interventions seldom provide substantial improvements~\citep{qian2024chatdevcommunicativeagentssoftware, wu2024stateflowenhancingllmtasksolving}. Furthermore, we tried evaluating Biomni/Phylo within the LLM judge suite, but its access to recently updated papers and external validation evidence may exceed the judges' closed-book biomedical knowledge,
particularly for claims grounded in post-cutoff literature. 

\section*{Impact Statement}

DeepRoot is a research tool for hypothesis generation, not medical advice.
By converting historical materia medica into auditable
source--compound--target--disease chains, it may help researchers prioritize
natural-product candidates for experimental follow-up and drug repurposing.

The main risks are overinterpretation, unsafe self-medication, and misuse of
traditional knowledge. Historical claims may be ineffective, toxic, or
culturally specific, and graph-supported plausibility does not establish
safety or efficacy. Any downstream use requires expert review, provenance
tracking, toxicity assessment, and experimental validation.

\nocite{langley00}

\section*{Data Availability}
The \emph{Shen Nong BenCao Jing} corpus and code for generating the KG and traversing it are provided at \href{https://github.com/CarlisleMa/deeprootv1}{\texttt{github.com/CarlisleMa/deeprootv1}}. Provided within are also the scripts and raw data generated for the evaluations.

\bibliography{references}
\bibliographystyle{icml2026}

\newpage
\appendix
\onecolumn
\section{Technical appendices and supplementary material}

\subsection{Knowledge graph schema}
\label{app:schema}

\paragraph{Edge types.} Seven typed edges, each carrying a numeric
\texttt{confidence\_score} (flat prior, see
Appendix~\ref{app:priors}), an \texttt{evidence\_type} tag
(see Appendix~\ref{app:graph-stats}), and a \texttt{source\_db}
provenance field where applicable.
\begin{itemize}
  \item \texttt{TREATS\_TRADITIONALLY} (Source $\to$ Traditional\_Malady):
        evidence span quoted from the source chunk text.
  \item \texttt{MAPS\_TO} (Traditional\_Malady $\to$ Modern\_Disease):
        \texttt{is\_primary}, \texttt{mapping\_role} $\in$
        \{primary, syndrome\_component\}, \texttt{mapping\_source}
        $\in$ \{gemini+icd10\_exact, gemini+mesh\_exact, gemini+snomed\_exact,
        gemini\_unverified\}, \texttt{mapping\_alternatives} (JSON).
  \item \texttt{IS\_EXTRACTED\_FROM} (Chemical\_Compound $\to$ Source):
        evidence type encodes COCONUT/PubChem provenance and resolution
        level (canonical vs.\ alias vs.\ formula).
  \item \texttt{TARGETS} (Chemical\_Compound $\to$ Biological\_Target):
        \texttt{pchembl\_score}, \texttt{assay\_id}, \texttt{assay\_type}
        $\in$ \{B, F\}, \texttt{assay\_description},
        \texttt{mechanism\_action}.
  \item \texttt{RELATES\_TO} (Biological\_Target $\to$ Modern\_Disease):
        \texttt{ot\_overall\_score}, \texttt{match\_tier} $\in$
        \{efo\_id, mondo\_id, doid\_id, mesh\_id, norm\_name\}.
  \item \texttt{KNOWN\_TREATS} (Chemical\_Compound $\to$ Modern\_Disease):
        \texttt{clinical\_phase} $\in \{1,2,3,4\}$, materialized from
        ChEMBL drug indications; held-out evaluation slice
        (\S\ref{sec:compound-recovery}).
  \item \texttt{PREPARED\_AS} (Source $\to$ Preparation\_Method).
\end{itemize}

\paragraph{Identity and write semantics.} All writes are idempotent
\texttt{MERGE}-on-identity. Compound identity is the RDKit-computed
InChIKey, which is invariant to canonical-SMILES variants and to
naming differences across COCONUT and PubChem. Target identity is
the ChEMBL ID, which unifies SINGLE PROTEIN, PROTEIN COMPLEX, PROTEIN
FAMILY, and ORGANISM target types under one key. Modern disease
identity is the canonical name, with ontology codes coalesce-backfilled
as later agents verify them against additional services.

\renewcommand{\thefigure}{S\arabic{figure}}
\setcounter{figure}{0}
\renewcommand{\thetable}{S\arabic{table}}
\setcounter{table}{0}

\begin{figure}[H]
    \centering
    \includegraphics[scale=0.5]{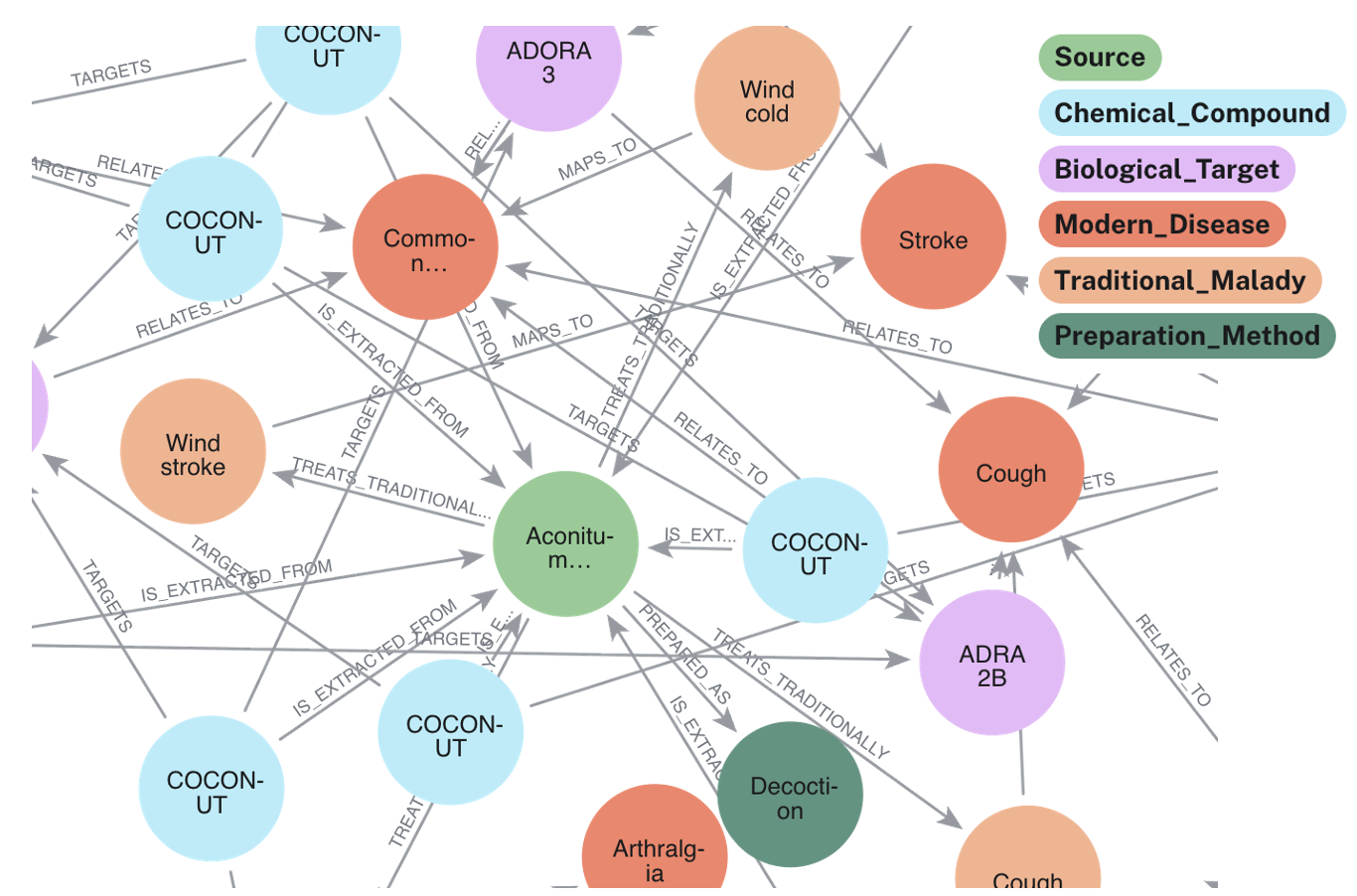}
    \caption{Example of a node cluster in Neo4j.}
    \label{fig:neo4j-example}
\end{figure}

\renewcommand{\arraystretch}{1.12}
\small
\setlength{\LTleft}{0pt}
\setlength{\LTright}{0pt}
\setlength{\tabcolsep}{3pt}

\begin{longtable}{@{}llp{7.2cm}@{}}
\caption{Complete node schema for the DeepRoot knowledge graph.}
\label{tab:kg-schema-full} \\
\toprule
\textbf{Node} & \textbf{Property} & \textbf{Description} \\
\midrule
\endfirsthead

\multicolumn{3}{c}{\tablename~\ref{tab:kg-schema-full} \textit{(continued)}} \\
\toprule
\textbf{Node} & \textbf{Property} & \textbf{Description} \\
\midrule
\endhead

\midrule
\multicolumn{3}{r}{\textit{Continued on next page}} \\
\endfoot

\bottomrule
\endlastfoot

\multirow{13}{*}{\texttt{Source}}
  & \texttt{name}                     & Canonical Latin binomial or common name (primary key) \\*
  & \texttt{aliases}                  & Alternative names from source text \\*
  & \texttt{evidence\_span}           & Verbatim passage from which the node was extracted \\*
  & \texttt{source\_document}         & Origin corpus file identifier \\*
  & \texttt{canonical\_name}          & Auditor-resolved canonical name \\*
  & \texttt{canonical\_type}          & Taxonomic category (herb, mineral, animal, fungus, \ldots) \\*
  & \texttt{canonical\_part}          & Plant/animal part used (root, bark, seed, whole, \ldots) \\*
  & \texttt{canonical\_source}        & Database used to resolve canonical form \\*
  & \texttt{canonical\_raw\_response} & Raw LLM response from canonicalization step \\*
  & \texttt{linker\_status}           & Compound-linker outcome (\texttt{ok}, \texttt{skipped}, \texttt{failed}) \\*
  & \texttt{linker\_attempted\_at}    & ISO timestamp of last linker run \\*
  & \texttt{linker\_compound\_count}  & Number of compounds linked from this source \\*
  & \texttt{linker\_evidence\_type}   & Evidence type used (\texttt{coconut}, \texttt{chembl}) \\
\midrule

\multirow{22}{*}{\texttt{Chemical\_Compound}}
  & \texttt{name}                        & IUPAC or common compound name (primary key) \\*
  & \texttt{smiles}                      & Canonical SMILES string \\*
  & \texttt{inchikey}                    & Standard InChIKey identifier \\*
  & \texttt{molecular\_formula}          & Molecular formula (e.g.\ \texttt{C21H23NO4}) \\*
  & \texttt{np\_likeness}                & Natural-product likeness score from COCONUT ($-5$ to $+5$) \\*
  & \texttt{annotation\_level}           & Structural confidence tier (1\,=\,MS2 confirmed, 5\,=\,predicted) \\*
  & \texttt{source\_db}                  & Source database (\texttt{COCONUT}, \texttt{ChEMBL}, \ldots) \\*
  & \texttt{coconut\_row}                & COCONUT row index for traceability \\*
  & \texttt{pubchem\_cid}                & PubChem Compound ID \\*
  & \texttt{target\_linker\_status}      & ChEMBL target-linker outcome \\*
  & \texttt{linker\_attempted\_at}       & ISO timestamp of target-linker run \\*
  & \texttt{linker\_chembl\_id}          & ChEMBL molecule ID used for target lookup \\*
  & \texttt{linker\_lookup\_method}      & Match method (\texttt{inchikey}, \texttt{smiles}, \texttt{name}) \\*
  & \texttt{linker\_target\_count}       & Targets written after filtering \\*
  & \texttt{linker\_dropped\_count}      & Targets dropped below pChEMBL floor \\*
  & \texttt{linker\_pchembl\_floor}      & pChEMBL activity threshold applied \\*
  & \texttt{linker\_max\_targets}        & Cap on targets written per compound \\*
  & \texttt{kt\_linker\_status}          & \texttt{KNOWN\_TREATS} linker outcome \\*
  & \texttt{kt\_linker\_attempted\_at}   & ISO timestamp of \texttt{KNOWN\_TREATS} linker run \\*
  & \texttt{kt\_linker\_indication\_count} & Drug indications written \\*
  & \texttt{kt\_linker\_dropped\_count}  & Indications dropped below phase threshold \\*
  & \texttt{kt\_linker\_min\_phase}      & Minimum clinical trial phase accepted \\
\midrule

\multirow{12}{*}{\texttt{Biological\_Target}}
  & \texttt{name}                        & Target protein name (primary key) \\*
  & \texttt{target\_pref\_name}          & ChEMBL preferred target name \\*
  & \texttt{gene\_symbol}                & HGNC gene symbol \\*
  & \texttt{uniprot\_id}                 & UniProt accession \\*
  & \texttt{target\_chembl\_id}          & ChEMBL target identifier \\*
  & \texttt{target\_type}                & Target class (SINGLE PROTEIN, PROTEIN COMPLEX, \ldots) \\*
  & \texttt{ncbi\_tax\_id}               & NCBI taxonomy ID of the target organism \\*
  & \texttt{td\_linker\_status}          & Target-disease linker outcome \\*
  & \texttt{td\_linker\_attempted\_at}   & ISO timestamp of target-disease linker run \\*
  & \texttt{td\_linker\_association\_count} & Disease associations written \\*
  & \texttt{td\_linker\_dropped\_count}  & Associations dropped below score threshold \\*
  & \texttt{td\_linker\_min\_score}      & Open Targets association score floor applied \\
\midrule

\multirow{8}{*}{\texttt{Modern\_Disease}}
  & \texttt{name}         & Disease name (primary key) \\*
  & \texttt{doid\_id}     & Disease Ontology identifier \\*
  & \texttt{mondo\_id}    & MONDO disease ontology identifier \\*
  & \texttt{mesh\_id}     & MeSH descriptor \\*
  & \texttt{efo\_id}      & Experimental Factor Ontology (EFO) identifier \\*
  & \texttt{icd10\_code}  & ICD-10 classification code \\*
  & \texttt{snomed\_id}   & SNOMED CT concept identifier \\*
  & \texttt{verified\_by} & Agent or curator that confirmed the mapping \\
\midrule

\multirow{11}{*}{\texttt{Traditional\_Malady}}
  & \texttt{name}                   & TCM ailment name (primary key) \\*
  & \texttt{description}            & Classical definition from source text \\*
  & \texttt{evidence\_span}         & Verbatim passage supporting the malady \\*
  & \texttt{source\_document}       & Origin corpus file identifier \\*
  & \texttt{mapper\_status}         & Malady-to-disease mapper outcome \\*
  & \texttt{mapper\_classification} & Ontology mapping confidence class \\*
  & \texttt{mapper\_attempted\_at}  & ISO timestamp of mapper run \\*
  & \texttt{mapper\_raw\_response}  & Raw LLM response from mapping step \\*
  & \texttt{archived}               & Reviewer flag: duplicate or low-quality node \\*
  & \texttt{archive\_reason}        & Free-text reason for archiving \\*
  & \texttt{reviewed\_by}           & Reviewer agent identifier \\
\midrule

\multirow{3}{*}{\texttt{Preparation\_Method}}
  & \texttt{name}           & Preparation name (decoction, pill, powder, \ldots) \\*
  & \texttt{route}          & Administration route (oral, topical, inhaled, \ldots) \\*
  & \texttt{evidence\_span} & Verbatim passage describing the preparation \\*
\end{longtable}

\renewcommand{\arraystretch}{1.0}
\normalsize


\subsection{Assembly agent protocols}
\label{app:agents}

\begin{table}[H]
\centering
\caption{The seven DeepRoot Assembly agents in dependency order.
Each agent combines LLM proposal with deterministic verification
against the listed grounding source.}
\label{tab:assembly-agents}

\footnotesize
\setlength{\tabcolsep}{2pt}
\renewcommand{\arraystretch}{0.95}

\resizebox{\columnwidth}{!}{%
\begin{tabular}{@{}c l p{4.8cm} p{2.2cm}@{}}
\toprule
\# & Agent & Role & Grounding source \\
\midrule

i
& Extraction
& emit Source / Malady / Preparation; \texttt{TREATS}, \texttt{PREPARED\_AS}
& --- \\

ii
& Auditor
& canonicalize; verify evidence spans; merge duplicates
& COCONUT, PubChem \\

iii
& Malady$\to$Disease
& generate-then-verify ontology mapping (\texttt{MAPS\_TO})
& MeSH, ICD-10, SNOMED \\

iv
& Source$\to$Compound
& natural-product / chemical lookup (\texttt{IS\_EXTRACTED\_FROM})
& COCONUT, PubChem \\

v
& Compound$\to$Target
& molecular targets and bioactivities (\texttt{TARGETS})
& ChEMBL \\

vi
& Target$\to$Disease
& mechanism-to-disease, dispatched on target type (\texttt{RELATES\_TO})
& Open Targets, NCBI Tax, OLS4 \\

vii
& Reviewer
& rules + LLM archival pass (orphans, OCR artifacts, off-domain entities)
& --- \\

\bottomrule
\end{tabular}%
}
\end{table}

\paragraph{Extraction.} A Gemini call per text chunk emits
\texttt{Source} nodes with aliases and species hints,
\texttt{Traditional\_Malady} nodes with descriptions,
\texttt{Preparation\_Method} nodes, and the
\texttt{TREATS\_TRADITIONALLY} and \texttt{PREPARED\_AS} edges among
them. Each edge records the literal evidence span from the source
chunk. Confidence is the LLM's self-assessed score, used only as a
soft signal for downstream auditing (the auditor verifies span and
identity independently).

\paragraph{Auditor.} Three deterministic post-extraction jobs.
(1)~\emph{Canonicalization.} Gemini Flash-Lite (temperature 0,
batched 20 sources/call, structured-JSON schema) labels each source
\texttt{organism} (Latin binomial + part), \texttt{chemical} (name or
formula), or \texttt{uncanonicalized}. Per-type external lookup
verifies: organisms hit a local COCONUT inverted index over
$62{,}792$ species keys, chemicals hit PubChem REST. Canonical
labels with external-DB hits are tagged \texttt{gemini+coconut} /
\texttt{gemini+pubchem}; the rest are tagged
\texttt{gemini\_unverified\_*}, \texttt{uncanonicalized}, or
\texttt{error}. (2)~\emph{Source merge.} Sources sharing the same
$(\texttt{canonical\_name}, \texttt{canonical\_part})$ collapse onto
a keeper (highest-degree, alphabetical tiebreak). Merged-from nodes
are soft-archived with reason \texttt{merged\_into:<keeper>};
outgoing edges are re-targeted (parallel edges take maximum
confidence) and aliases are unioned.
(3)~\emph{Evidence-span verification.} Substring check
(whitespace-normalized) of every \texttt{TREATS\_TRADITIONALLY}
evidence span against its source chunk. Hallucinated spans
(typically LLM-introduced ellipsis) trigger soft-archival with
reason \texttt{hallucinated\_evidence}.

\paragraph{Malady$\to$Disease (generate-then-verify).} One Gemini
call per malady (temperature 0, six-shot system prompt) emits a
typed exit: \texttt{disease}, \texttt{symptom}, \texttt{syndrome},
\texttt{ambiguous}, or \texttt{tcm\_no\_equivalent}. Crucially, the
LLM emits canonical \emph{names only}---never codes---together with
an ontology hint. Each proposed name is then verified in parallel
against ICD-10 (NLM Clinical Tables), MeSH (NLM RDF Lookup), and
SNOMED (EBI OLS4). Verification accepts only \emph{tolerant exact
match} (case- and punctuation-insensitive); fuzzy hits, even from
the API itself, are rejected. Syndromes can produce one primary
plus up to two \texttt{syndrome\_component} edges, but only
components that pass exact-match verification are written
(unverified components are dropped and logged in the primary edge's
\texttt{mapping\_alternatives}). Default mode rejects unverified
mappings entirely; an \texttt{--allow-unverified} flag stores them
with \texttt{requires\_review=true} and a degraded prior.

\paragraph{Source$\to$Compound.} Routes by \texttt{canonical\_type}.
Organisms hit a local COCONUT inverted index (in-memory,
$\sim$$700{,}000$ structures with species provenance) by exact
normalized species name and the ``first two tokens'' (genus species)
prefix. Chemicals hit a disk-cached PubChem REST client (4 RPS),
with formula fallback via \texttt{compound/fastformula} for
formula-shaped canonicals. Compound identity is the RDKit-computed
InChIKey from canonical SMILES, which collapses canonicalization
variants and unifies COCONUT/PubChem name splits onto a single
node. Edges are written with \texttt{evidence\_type} encoding the
resolution path (\texttt{coconut\_organism\_canonical},
\texttt{pubchem\_chemical\_canonical}, alias, formula, unverified).
Per-source \texttt{linker\_status} stamping makes the agent fully
resumable across network interruptions.

\paragraph{Compound$\to$Target.} ChEMBL queries by InChIKey
($91.4\%$ resolution rate), falling back to canonical SMILES then
preferred name. For each resolved compound, the agent retrieves
mechanism-of-action records and bioactivity records. Activity
filtering: \texttt{pchembl\_value} $\geq 5.0$ (10\,$\mu$M floor) for
quantitative tier, \texttt{assay\_type} $\in \{B, F\}$,
\texttt{standard\_relation} $\in \{=, \sim\}$; data validity flags
must be empty. Target type is \emph{not} restricted to SINGLE
PROTEIN: PROTEIN COMPLEX (subunit fan-out), PROTEIN FAMILY
(broad-spectrum inhibitors), and ORGANISM (anti-pathogen evidence,
e.g.\ \emph{Plasmodium falciparum} for antimalarials) are all
admitted. A \texttt{--include-phenotypic} flag additionally
retrieves phenotypic activities (\texttt{pchembl\_value} null,
\texttt{assay\_type} F/B) at a lower confidence prior; for terpenes,
sterols, and other natural products tested phenotypically rather
than against named molecular targets, this opt-in is necessary to
avoid silent loss of $\sim$$2{,}000$ compounds. Salt and tautomer
parents are aggregated via the ChEMBL molecule hierarchy; without
hierarchy expansion, $30$--$50\%$ of activities are missed for
multi-form compounds.

\paragraph{Target$\to$Disease.} Four-way dispatch on
\texttt{target\_type}. SINGLE PROTEIN: Open Targets GraphQL via
UniProt$\to$Ensembl, returning disease associations with overall
scores binned to confidence tiers. PROTEIN COMPLEX: subunit fan-out
via ChEMBL, then per-subunit Open Targets, with per-disease
max-score deduplication. PROTEIN FAMILY: intentionally skipped
(\texttt{td\_linker\_status = skipped\_protein\_family}), as
family-level evidence is too coarse for clinical association.
ORGANISM: an EFO/DOID walk over OLS4 starting from the NCBI
\texttt{tax\_id}, emitting both the specific disease class and its
ancestors, falling back to a parallelized LLM safety net (Gemini
proposes candidate disease names, NLM MeSH verifies via tolerant
exact match) when the ontology walk returns empty. Three-tier
disease matching: exact ontology-ID match first, then
normalized-name exact match, then MeSH-synonym expansion. A
plan-then-apply phase performs all writes via
\texttt{UNWIND}-batched Cypher transactions ($\sim$$12$ transactions
for $\sim$$5{,}500$ rows), with \texttt{DELETE} restricted to
terminal-status rows so transient API failures cannot wipe live
edges.

\paragraph{Reviewer.} Two-pass deterministic-then-LLM archival.
Pass~1 catches OCR artifacts (single-character entities, mojibake),
orphans (degree~$0$), generic categories (``herb'', ``compound''),
and metaphysical concepts that escaped extraction. Pass~2 batches
the residual ambiguous nodes ($\sim$$20$) to Gemini for
biomedical-relevance filtering. Cascade archival propagates to
incident edges. Archival is soft (\texttt{archived=true} with
reason); no records are deleted.


\subsection{Discovery agent protocols}
\label{app:discovery}

DeepRoot Discovery comprises two agent roles operating over the
typed graph: a \emph{critic agent} that scores existing
Source$\to$Malady claims, and a \emph{discovery agent} that nominates
novel compound candidates for a target Modern\_Disease. Both
consume the tier-bucketed path-scoring layer described in
Appendix~\ref{app:scoring}.

\paragraph{Critic agent.} For each (Source, Malady) claim, the
agent receives a structured payload assembled from the KG: the
claim itself (Source name + aliases, Malady description, primary
mapped Modern\_Disease, mapping rationale), deterministic Pass-1
signals (path bucket distribution, loop-closure counts, top-bucket
score), the top-$K$ mechanistic chains
(Source$\to$Compound$\to$Target$\to$Disease) with edge metadata, and
four cross-cutting enrichments---compound profiles
(\texttt{KNOWN\_TREATS} for other diseases, target spectrum), target
genericity (number of associated diseases per target), source-level
target convergence (multi-compound hits on the same target), and
sibling verdicts (Pass-1 verdicts of other claims on the same
source). The model returns a structured JSON
\texttt{CriticVerdict} with: a verdict on the four-tier ladder
(\textsc{validated} / \textsc{plausible} / \textsc{weak} /
\textsc{unsupported}); biological\_plausibility and
evidence\_coherence scores in $[0, 1]$, defensively clamped to that
range; a \texttt{key\_evidence} list of cited compound--target--
disease triples; a \texttt{concerns} list with typed enum values
(\texttt{generic\_target}, \texttt{weak\_evidence\_only},
\texttt{indirect\_mechanism}, \texttt{wrong\_disease\_mapping},
\texttt{syndrome\_underutilized},
\texttt{promiscuous\_compound},
\texttt{unverified\_evidence}); a free-form rationale; and a
\texttt{requires\_human\_review} flag (auto-set when the LLM and
deterministic Pass-1 verdicts disagree by $\geq 2$ rungs). The
prompt instructs the model to quote specific input fields and never
speculate beyond the provided evidence; numeric ranges are enforced
via post-hoc clamping rather than relying on the model to obey them.

\paragraph{Discovery (nominator) agent.} Given a target
Modern\_Disease query $d^*$, the agent walks the KG backward
(disease $\leftarrow$ malady $\leftarrow$ source $\leftarrow$
compound) to enumerate all corpus-supported candidate compounds,
then walks forward (compound $\to$ target $\to$ disease) to score
each candidate's mechanistic plausibility. A novelty filter drops
compounds whose \texttt{KNOWN\_TREATS} edge already reaches $d^*$
(supporting an in-memory mask for held-out evaluation
without mutating the graph). The remaining candidates are ranked
lexicographically by (i)~\texttt{has\_loop\_closure} (does at least
one forward chain reach $d^*$), (ii)~\texttt{forward\_bucket} (T1
$>$ T2 $>$ T3 $>$ T4; weakest-link tier of the strongest
loop-closing path), (iii)~\texttt{unique\_sources\_count},
(iv)~\texttt{unique\_maladies\_count}, and
(v)~\texttt{forward\_max\_score} (multiplicative product of
edge confidences along the strongest path). The output is an
ordered list of \texttt{CompoundCard} entries containing top
historical paths (source, malady, evidence span), top forward paths
(target, assay description, OT score), and KNOWN\_TREATS for other
diseases as polypharmacology context. The discovery agent is
fully deterministic---no LLM is in the loop---making the ranking
auditable and stable across re-runs.


\subsection{Tier-bucket path scoring}
\label{app:scoring}

A \emph{path} is a sequence of typed edges connecting a source node
to a disease node through compound and target intermediaries. Each
edge carries an \texttt{evidence\_type} tag (e.g.,
\texttt{chembl\_mechanism}, \texttt{ot\_association\_strong},
\texttt{coconut\_organism\_canonical}) which maps to one of four
tiers \(T \in \{\text{T1}, \text{T2}, \text{T3},
\text{T4}\}\) (Appendix~\ref{app:priors}, Table~\ref{tab:priors})
and a flat numeric prior \(c \in [0, 1]\).

For a path \(p\) with edges \(e_1, \ldots, e_n\), the
\emph{path bucket} \(B(p)\) and \emph{path score} \(S(p)\) are
\[
  B(p) \;=\; \min_{i=1\ldots n} T(e_i),
  \qquad
  S(p) \;=\; \prod_{i=1\ldots n} c(e_i).
\]
Paths are ordered lexicographically by \((B(p), S(p))\) with the
bucket as the primary key (highest-tier bucket wins) and the
multiplicative score as tiebreak within a bucket. The bucket
captures the qualitative claim ``a chain is only as strong as its
weakest edge'' (weakest-link), while the score gives a continuous
ordering inside each tier.

The same scoring layer is consumed by both Discovery agents
(Appendix~\ref{app:discovery}) and by the deterministic Pass-1
signals fed to the critic. Because priors are flat (Table~\ref{tab:priors})
rather than learned, raw external scores
(\texttt{ot\_overall\_score}, \texttt{pchembl\_value},
\texttt{np\_likeness}) are preserved as edge attributes so
downstream consumers can recalibrate without re-running Assembly.


\subsection{Confidence priors}
\label{app:priors}

\paragraph{Tier ladder (used for path scoring).} T1 $>$ T2 $>$
T3 $>$ T4. Path bucket is the minimum tier across edges
(weakest-link); within a bucket, ranking uses the multiplicative
product of edge confidences as tiebreak.

\begin{table}[H]
\centering
\caption{Per-edge-type confidence priors. Priors are flat (not
learned), chosen on biomedical face validity, and never replaced by
self-reported LLM confidence. Raw external scores (e.g.,
\texttt{ot\_overall\_score}, \texttt{pchembl\_value},
\texttt{np\_likeness}) are preserved as edge properties so
downstream consumers can recalibrate without re-running Assembly.}
\label{tab:priors}
\footnotesize
\setlength{\tabcolsep}{3pt}
\resizebox{\linewidth}{!}{%
\begin{tabular}{@{}l >{\raggedright\arraybackslash}p{6.5cm} l@{}}
\toprule
Edge & Evidence type & Tier (prior) \\
\midrule
\multirow{4}{*}{\texttt{IS\_EXTRACTED\_FROM}}
  & coconut\_organism\_canonical / pubchem\_chemical\_canonical    & T1 (0.70--0.80) \\
  & coconut\_organism\_alias                                       & T2 (0.55) \\
  & coconut\_organism\_unverified / pubchem\_chemical\_unverified  & T3 (0.50--0.55) \\
  & pubchem\_chemical\_formula                                     & T4 (0.50) \\
\midrule
\multirow{4}{*}{\texttt{TARGETS}}
  & chembl\_mechanism                                              & T1 (0.95) \\
  & chembl\_activity\_strong (pchembl $\geq 7$)                    & T2 (0.75) \\
  & chembl\_activity\_moderate (pchembl $\geq 6$)                  & T3 (0.60) \\
  & chembl\_activity\_weak (pchembl $\geq 5$) / chembl\_phenotypic & T4 (0.40) \\
\midrule
\multirow{4}{*}{\texttt{RELATES\_TO}}
  & ncbi\_pathogen\_consensus                                      & T1 (0.92) \\
  & ot\_association\_strong (OT $\geq 0.7$)                        & T1 (0.85) \\
  & ot\_association\_moderate (OT $\geq 0.4$) / complex\_aggregate / pathogen\_llm\_verified & T2 (0.65--0.75) \\
  & ot\_association\_weak (OT $\geq 0.2$)                          & T3 (0.45) \\
\midrule
\multirow{2}{*}{\texttt{MAPS\_TO}}
  & icd10/mesh/snomed exact (primary)                              & T1 (0.80--0.85) \\
  & syndrome\_component / symptom                                  & T2 (0.65--0.75) \\
\midrule
\multirow{4}{*}{\texttt{KNOWN\_TREATS}}
  & clinical\_phase = 4 (approved)                                 & T1 (0.95) \\
  & clinical\_phase = 3                                            & T1 (0.85) \\
  & clinical\_phase = 2                                            & T2 (0.65) \\
  & clinical\_phase = 1                                            & T3 (0.45) \\
\bottomrule
\end{tabular}%
}
\end{table}


\subsection{Graph statistics}
\label{app:graph-stats}

\paragraph{Node and edge totals (active, post-Assembly).} $21{,}111$
nodes active, $94$ archived. By type: $415$ Source, $294$
Traditional\_Malady, $129$ Modern\_Disease, $18{,}012$
Chemical\_Compound, $2{,}211$ Biological\_Target, $50$
Preparation\_Method. Edges: $52{,}467$ active. By type: $32{,}909$
\texttt{IS\_EXTRACTED\_FROM}, $16{,}696$ \texttt{TARGETS}, $1{,}841$
\texttt{RELATES\_TO}, $431$ \texttt{TREATS\_TRADITIONALLY}, $301$
\texttt{KNOWN\_TREATS}, $257$ \texttt{MAPS\_TO} ($208$ primary $+
49$ syndrome\_component), $32$ \texttt{PREPARED\_AS}.

\paragraph{Per-evidence-type breakdown.}
\texttt{IS\_EXTRACTED\_FROM}: $32{,}885$ organism\_canonical, $21$
chemical\_canonical, $3$ formula. \texttt{TARGETS}: $60$ mechanism,
$1{,}148$ strong, $1{,}185$ moderate, $2{,}264$ weak, $12{,}039$
phenotypic. \texttt{RELATES\_TO}: $936$ ot\_weak, $666$
ot\_moderate, $14$ ot\_strong, $203$ complex\_aggregate, $17$
llm\_verified, $3$ pathogen\_consensus, $2$ efo.
\texttt{KNOWN\_TREATS}: $60$ phase~$4$, $92$ phase~$3$, $84$
phase~$2$, $65$ phase~$1$. The phenotypic-heavy distribution of
\texttt{TARGETS} reflects the natural-product corpus: terpenes,
sterols, and flavonoids are predominantly characterized by
phenotypic bioassays rather than named molecular targets.

\paragraph{Convergence.} $4{,}605$ compounds appear in $\geq 2$
sources (classic phytomedicine pattern: $\beta$-sitosterol
$112\times$, quercetin $87\times$, kaempferol $66\times$). Across
the $129$ \texttt{Modern\_Disease} nodes, $257$ \texttt{MAPS\_TO}
edges resolve to an average of $1.99$ maladies per disease,
indicating strong canonical convergence rather than fragmentation.

\paragraph{Coverage.} $504$ of $2{,}211$ targets ($22.8\%$) link to
at least one disease; $88$ of $129$ disease nodes ($68\%$) are
reached by at least one target. $3{,}221$ of $18{,}012$ compounds
($17.9\%$) have at least one \texttt{TARGETS} edge; the remainder
either lack ChEMBL records or have no admissible target-class data,
a documented limitation of the underlying databases rather than of
the pipeline.


\subsection{Evaluation protocols}
\label{app:eval-protocols}

\paragraph{Eval~1: edge-perturbation sensitivity.} A fixed test set of closed-loop Source--Malady claims is sampled from the KG.
For each perturbation level $p \in \{0\%, 20\%, 40\%, 60\%, 80\%, 100\%\}$,
a fraction $p$ of edges across four mechanistic edge types
(\texttt{TARGETS}, \texttt{RELATES\_TO}, \texttt{KNOWN\_TREATS}, \texttt{MAPS\_TO})
is selected uniformly at random and their target endpoints are shuffled among
themselves. A single perturbation is
applied per level; all test claims are then evaluated by the Critic against the
same perturbed graph. Self-confidence (mean biological plausibility over all
claims) is reported per level.

\paragraph{Eval~2: source and compound recovery on mini-corpora.}
To build each mini-corpus, \texttt{build\_recovery\_eval\_corpus.py} first
queries the knowledge graph for two disjoint source pools:
\textbf{closed-loop sources} (those with at least one complete
\texttt{Source $\to$ Compound $\to$ Target $\to$ Disease} chain where the
source also \texttt{TREATS\_TRADITIONALLY} $\to$ \texttt{Malady} $\to$
\texttt{MAPS\_TO} the same disease) and \textbf{distractor sources} (all other
non-archived KG sources). The full \textit{Shen Nong Ben Cao Jing} text is
split into paragraphs, each tagged by keyword regex against every source name
in the KG. Gemini Flash then verifies which tagged sources a paragraph actually
\emph{describes therapeutically} (rather than merely cross-referencing in a
compatibility list). Verified single-source paragraphs are banked into the two
pools. Each synthetic eval case is assembled by a diversity-maximising greedy
algorithm: it picks $K{=}3$ least-used closed-loop source paragraphs and
$N{=}7$ least-used distractor source paragraphs, ensures no overlap between
the two sets, then deterministically shuffles all 10 paragraphs into an
interleaved mini-corpus. The label set for compound recovery includes both
\textbf{closed-loop compounds} (retrieved via a Cypher walk confirming
\texttt{Compound $\to$ TARGETS $\to$ Target $\to$ RELATES\_TO $\to$ Disease}
for a disease the source already treats; typically 1--5 per source) and
\textbf{distractor compounds} (\texttt{IS\_EXTRACTED\_FROM} compounds of the
distractor sources). Compound recall@$k$ is reported separately against each
label set so the closed-loop and broad-coverage signals can be distinguished.
Thirty such mini-corpora are generated, each with a distinct closed-loop source
signature enforced by deduplication.

\paragraph{Eval~3: positive-control recovery of hidden known
treatments.} The 301 \texttt{KNOWN\_TREATS} edges are filtered to
those whose Modern\_Disease has a backward chain
$d \leftarrow$ malady $\leftarrow$ source $\leftarrow$ compound in
the KG (the \emph{historical-reachability} subset), yielding
21 (compound, disease) pairs across 10 diseases. For each test
pair $(c^*, d^*)$, we compute $c^*$'s planar InChIKey prefix (first
14 characters, dropping stereochemistry) and mask every
\texttt{KNOWN\_TREATS} edge from any compound sharing that prefix to
$d^*$ (in-memory mask only; the graph is not mutated). The
discovery agent is then run on $d^*$ with top-$K = 20$. A trial
succeeds at rank $r$ if any nominee within the top $r$ shares
$c^*$'s planar prefix. The candidate pool---all compounds
reachable via the backward chain from $d^*$---is recorded
per-disease (range 87--1{,}954, median 835), giving a uniform
random recall@$20$ baseline of $\approx 2.4\%$.

\paragraph{Eval~4: LLM-as-judge reasoning quality.} A stratified
sample of 30 closed-loop Source--Malady claims is drawn from the
431 candidate claims, with the strata chosen to exercise distinct
verdict regimes (representative balance of \emph{unsupported},
\emph{strong\_support}, \emph{T1-bucket-without-loop-closure},
\emph{mechanistic-only}, and \emph{traditional-only}). The same 30
claims are scored by six conditions: DeepRoot Discovery at three
LLM tiers (Gemini~3.1~Pro, 2.5~Flash, 3.1~Flash-Lite), a
graph-only deterministic baseline (Pass-1 verdict only, no LLM),
an LLM baseline given just the corpus passages, and a tool-call
LLM baseline given direct API access to ChEMBL, Open Targets,
PubMed, and MeSH. Outputs are graded by Claude Sonnet 4.6
(cross-family from the graded systems) on six dimensions in
$[1, 5]$: Evidence Fidelity (does the critic cite evidence present
in the payload?), Verdict Alignment (does the verdict follow from
the visible evidence?), Reasoning Coherence (does the rationale
explain \emph{this} claim's chain?), Clinical Mapping (does the
critic responsibly handle the malady$\to$disease mapping?),
Uncertainty Calibration (are the scores, concerns, and review
flag calibrated?), and Actionability (would a curator know what to
inspect next?). The judge additionally returns six binary flags
(\texttt{hallucinated\_evidence},
\texttt{unsupported\_verdict\_jump},
\texttt{ignored\_loop\_closure\_status},
\texttt{overclaims\_strength},
\texttt{contradictory\_scores},
\texttt{needs\_human\_review}) and a recommended status in
\{\texttt{pass}, \texttt{weak\_pass}, \texttt{fail},
\texttt{human\_review}\}. The judge sees only the critic's visible
artifacts (verdict, scores, key\_evidence, concerns, rationale)
plus, where applicable, the structured payload that the critic was
given; it does not have access to ground truth and grades the
quality of the critic's argument rather than its absolute
correctness.


\subsection{Prompt templates}
\label{app:prompts}

This section summarizes the four critical prompts in DeepRoot.
Each template is described as a tuple of
\emph{(role, input schema, output schema, key instructions)}; the
verbatim text is in the released codebase.

\paragraph{Extraction prompt.} \emph{Role:} extract typed entities
from a single text chunk of the source corpus. \emph{Input:} the
chunk text plus a \texttt{source\_document} identifier. \emph{Output
schema (structured JSON):} \texttt{sources[]},
\texttt{maladies[]}, \texttt{preparations[]},
\texttt{treats\_edges[]}, \texttt{prepared\_as\_edges[]}; each
entity carries \texttt{name}, \texttt{aliases},
\texttt{evidence\_span} (verbatim quote from the chunk), and a
self-assessed \texttt{confidence}. \emph{Key instructions:}
evidence spans must be substring-matchable to the chunk text;
identifiers (binomials, ChEMBL IDs) are never to be invented; the
LLM emits names, not codes.

\paragraph{Malady$\to$Disease mapping prompt
(generate-then-verify).} \emph{Role:} classify each
Traditional\_Malady into one of \{\texttt{disease},
\texttt{symptom}, \texttt{syndrome}, \texttt{ambiguous},
\texttt{tcm\_no\_equivalent}\} and propose canonical English
name(s). \emph{Input:} the malady's \texttt{name},
\texttt{description}, \texttt{evidence\_span}, and source
classical context. \emph{Output schema:} a top-level
\texttt{classification} field plus a \texttt{mappings[]} list, where
each mapping carries \texttt{name} (canonical English), an
\texttt{ontology\_hint} $\in \{\texttt{icd10}, \texttt{mesh},
\texttt{snomed}\}$, a \texttt{role}
$\in \{\texttt{primary}, \texttt{syndrome\_component}\}$, and a
one-sentence \texttt{rationale}. \emph{Key instructions:}
the model emits canonical \emph{names only, never codes}; for
\texttt{syndrome} maladies, up to two
\texttt{syndrome\_component} entries may be returned in addition to
the primary; six in-context examples cover the typical TCM
patterns (\textit{wind heat}, \textit{gu toxin},
\textit{counterflow}, etc.). Codes (ICD-10, MeSH, SNOMED) are
recovered downstream by deterministic exact-match verification
against the corresponding ontology services---never trusted from
the LLM.

\paragraph{Critic agent prompt.} \emph{Role:} biomedical reasoning
expert evaluating whether a historical Source plausibly treats a
Modern\_Disease via known mechanisms. \emph{Input:} the structured
payload described in Appendix~\ref{app:discovery} (claim,
Pass-1 signals, evidence paths, compound profiles, target
profiles, source-level target convergence, sibling verdicts),
serialized as a single JSON object. \emph{Output schema:} the
\texttt{CriticVerdict} object described in
Appendix~\ref{app:discovery}. \emph{Key instructions
(highlighted in the system prompt):} (i)~assess
\textsc{target quality} via the per-target disease count
(pleiotropy of $\geq 50$ associated diseases is flagged
\texttt{generic\_target}); (ii)~ground
\textsc{compound pharmacology} in the compound's
\texttt{KNOWN\_TREATS} record and target spectrum;
(iii)~scrutinize the
\textsc{disease mapping} for clinical plausibility against the
historical evidence span; (iv)~upweight
\textsc{polypharmacology / convergence} when multiple compounds
from the same source hit the same target;
(v)~discount \textsc{specificity} when the source has many
unrelated claims (kitchen-sink remedy). The prompt instructs the
model to quote specific input fields and never speculate beyond
the provided payload.

\paragraph{Baseline LLM prompt.} \emph{Role:} Candidate nominator. You are a pharmaceutical and natural products expert with broad knowledge of traditional Chinese medicine, pharmacognosy, and bioactive plant compounds. You will be given a passage from the Shen Nong Ben Cao Jing and will be asked to discover and rank plausibility of therapeutic compounds. Read this passage from a historical Chinese herbal text. List each medicinal source discussed; for each, propose up to 10 therapeutic chemical compounds it likely contains, with a plausibility score 0.0-1.0 (0.0 = incoherent or no known mechanism, 1.0 = biologically obvious) and a one-sentence reasoning grounded in known biochemistry or pharmacology. 

\paragraph{Biomni prompt.} \emph{Role:} Biomedical reasoning expert. \emph{Input:} You are a pharmaceutical and natural-products expert with broad knowledge of
traditional Chinese medicine, pharmacognosy, bioactive plant/mineral/animal
compounds, and modern clinical pharmacology. You will be given an extracted corpus of historical medicinal entries. Each
block is one claim: a SOURCE (a medicinal substance) and a MALADY (the
traditional ailment the historical text says it treats), followed by the
original passage. You must judge to your fullest capability, using any databases and subagents available necessitating @ChemBL, @PubChem, @PubMed, @OpenTargets as places for you to find relevant compounds or links to mechanistically evaluate the plausibility that the source provided can treat the malady listed.  
Rules: Score each claim independently; do not let one claim bias another. Do not invent citations or database IDs. Reason from known pharmacology. If the source is an inert mineral or has no plausible bioactive route to the disease, say so and mark Unsupported.

\paragraph{LLM-judge prompt.} \emph{Role:} independent biomedical
evaluation judge grading the visible reasoning of an automated
critic, not deciding whether the underlying therapeutic claim is
true. \emph{Input:} \texttt{condition} under judgment (one of the
six in Eval~4), the claim, the Pass-1 deterministic signals, the
visible payload the critic received (or empty for graph-free
conditions), and the critic's full structured output.
\emph{Output schema:} per-dimension scores in $[1, 5]$ for the six
rubric dimensions (Appendix~\ref{app:eval-protocols}), six binary
flags, a recommended status, and a free-text justification
referencing specific input fields. \emph{Key instructions:}
treat Pass-1 as an input signal, not ground truth (a critic can be
good even when it agrees with an imperfect Pass-1 verdict, if it
explains the limitation correctly); penalize hallucinated
citations (every compound, target, and disease named by the critic
must trace to a payload field, except for widely accepted
biomedical facts in text-only conditions); penalize rationales
that confuse global top-bucket evidence with loop-closing
disease support (a claim with 76 T1-bucket paths but zero of
those paths reaching the mapped disease is \emph{not}
well-supported).


\subsection{Implementation notes}
\label{app:impl}

\paragraph{Models.} Extraction, Auditor canonicalization, Malady
mapper, and Reviewer Pass~2 use Gemini~3.1-Flash-Lite at temperature 0
with structured-JSON schemas (rate-limited 1 request/sec, retried
up to 4 times on transient failure). The Discovery LLM stages use
Gemini~Pro under the same rate-limit envelope.

\paragraph{Storage.} Neo4j AuraDB. All writes are idempotent
\texttt{MERGE} statements with status fields (e.g.,
\texttt{linker\_status}, \texttt{td\_linker\_status},
\texttt{mapper\_status}) written \emph{last} so transient failures
retry safely. Default re-run filters select only nodes with
\texttt{status IS NULL}; \texttt{--retry-misses} reprocesses
\texttt{error} and \texttt{no\_*} branches; \texttt{--force-relink}
deletes prior edges from terminal-status branches only, never from
in-flight nodes. This makes the pipeline crash-safe end-to-end.

\paragraph{Caching.} PubChem responses are disk-cached by query;
COCONUT is loaded once into an in-memory inverted index. Re-runs
after network interruptions complete in seconds.

\paragraph{End-to-end cost.} Full Assembly over the 71-chunk corpus
runs in $\sim$$30$--$40$ minutes wall-clock, dominated by per-edge
Neo4j MERGE latency on AuraDB. Total LLM spend is $\sim$\$$0.25$
(Flash-Lite); ChEMBL, COCONUT, PubChem, Open Targets, NLM, and EBI
are all free.
\clearpage


\subsection{Qualitative scoring of representative critic agent responses}
\label{app:examples}

We assembled fifty source--malady pairs, constrained so that every pair has both a traditional \texttt{TREATS\_TRADITIONALLY} edge and a primary mapping to a modern disease. For each pair we extracted the corresponding historical passage from the \textit{Shen Nong Ben Cao Jing} to form a mini-corpus. The mini-corpus was given to DeepRoot Discovery, Biomni, or baseline LLM and tasked to score each pair based on the plausibility that the particular source can treat the traditional malady it is claimed to treat. Shown below are the first 10 examples for DeepRoot and Biomni.

\begin{table}[H]
  \caption{DeepRoot versus Biomni
  baseline): malady$\to$modern-disease mappings and plausibility verdicts
  across ten (Source, Malady) pairs.}
  \label{tab:deeproot-vs-biomni}
  \centering
  \footnotesize
  \setlength{\tabcolsep}{3pt}
  \begin{tabular}{@{}c >{\raggedright\arraybackslash}p{3.0cm} >{\raggedright\arraybackslash}p{2.3cm} >{\raggedright\arraybackslash}p{1.7cm} >{\raggedright\arraybackslash}p{2.5cm} >{\raggedright\arraybackslash}p{1.7cm}@{}}
    \toprule
    & & \multicolumn{2}{c}{DeepRoot} & \multicolumn{2}{c}{Biomni} \\
    \cmidrule(lr){3-4}\cmidrule(lr){5-6}
    \# & Source $\to$ Malady & Mapped disease & Verdict & Mapped disease & Verdict \\
    \midrule
    1 & \textit{Epimedium} $\to$ impotence & Erectile Dysfunction & Very Plausible & Erectile Dysfunction & Very Plausible \\
    2 & \textit{Trichosanthes cucumeroides} $\to$ menstrual block & Amenorrhea & Unsupported & Amenorrhea & Plausible \\
    3 & \textit{Rhizoma Arisaematis} $\to$ heart pain & Angina Pectoris & Previously Reported & Angina Pectoris & Weak \\
    4 & \textit{Calcareous Spar} $\to$ generalized fever & Fever & Unsupported & Pyrexia (Fever) & Unsupported \\
    5 & \textit{Aster tataricus} $\to$ cough & Cough & Plausible & Bronchitis; Asthma & Very Plausible \\
    6 & \textit{Fructus Ailanthi Altissimi} $\to$ impotence & Erectile Dysfunction & Plausible & Erectile Dysfunction & Weak \\
    7 & \textit{Ge Gen} $\to$ toxicity & Poisoning & Plausible & Alcohol Poisoning & Plausible \\
    8 & \textit{Sanguisorba officinalis} $\to$ vaginal discharge & Leukorrhea & Unsupported & Leukorrhea; Candidiasis & Plausible \\
    9 & \textit{Hedgehog Pelt} $\to$ hemorrhoids & Hemorrhoids & Plausible & Hemorrhoidal Disease & Unsupported \\
    10 & \textit{Calcareous Spar} $\to$ cough & Cough & Unsupported & Cough; Bronchitis & Unsupported \\
    \bottomrule
  \end{tabular}
\end{table}

{\footnotesize
Each box below shows one verdict for a (Source, Malady) pair.
\verdictbadge{verifiedgreen}{VERY PLAUSIBLE} \verdictbadge{verifiedgreen}{PREVIOUSLY REPORTED}
\verdictbadge{plausibleblue}{PLAUSIBLE} \verdictbadge{weakorange}{WEAK}
\verdictbadge{unsupportedred}{UNSUPPORTED}
badges reflect the reported plausibility tier (green = strongest; \textsc{Previously Reported}
denotes a source compound with an already-known therapeutic edge to the mapped disease).
}

{\small\bfseries\color{sectionblue}Example 1: \textit{Epimedium} (Source) and \textit{Impotence} (Malady)}
\vspace{2pt}\color{midgray}\hrule\color{black}\vspace{4pt}

\begin{mdframed}[backgroundcolor=lightgray,linecolor=midgray,roundcorner=4pt,
    innertopmargin=8pt,innerbottommargin=8pt]
  \textbf{DeepRoot Verdict:}\quad \verdictbadge{verifiedgreen}{VERY PLAUSIBLE}

  \textbf{Weakest Link:} The KG contains several non-closing paths for icariin that distract from the primary, well-established PDE5 inhibitory mechanism.

  \textbf{Reasoning:} Epimedium is well-documented for its primary active constituent, icariin, a known phosphodiesterase-5 (PDE5) inhibitor. The KG confirms the icariin$\to$PDE5A$\to$erectile dysfunction path, which is the canonical mechanism for treating impotence by enhancing nitric oxide-mediated vasodilation. Although other paths in the graph do not close the loop, the primary mechanism is highly robust and clinically validated.

  \vspace{3pt}

  \vspace{6pt}\color{midgray}\hrule\color{black}\vspace{4pt}

  \textbf{Biomni Verdict:}\quad \verdictbadge{verifiedgreen}{VERY PLAUSIBLE}

  \textbf{Weakest Link:} Icariin's PDE5A IC$_{50}$ ($\sim$5,900 nM) is $\sim$1,000-fold weaker than sildenafil ($\sim$3.5 nM); oral bioavailability of the native glycoside in humans remains the key uncertainty.

  \textbf{Reasoning:} Icariin (CID 5318997; CHEMBL553204) is a confirmed PDE5A inhibitor (IC$_{50}$ = 5,900 nM, ChEMBL; active in PubChem AID 362785 --- human recombinant PDE5A1 assay), sharing the same mechanism as sildenafil: cGMP preservation $\to$ corpus cavernosum smooth muscle relaxation. In vivo rat models confirm efficacy via GRK2 downregulation in penile tissue (PMID 39905744) and protection of cavernous tissue under hyperglycemia (PMID 40161546). The metabolite icariside II is significantly more potent. Icariin has reached ChEMBL Max Phase 3, making this the best-validated claim in the dataset.
\end{mdframed}

\vspace{6pt}
{\small\bfseries\color{sectionblue}Example 2: \textit{Trichosanthes cucumeroides} (Source) and \textit{Menstrual block} (Malady)}
\vspace{2pt}\color{midgray}\hrule\color{black}\vspace{4pt}

\begin{mdframed}[backgroundcolor=lightgray,linecolor=midgray,roundcorner=4pt,
    innertopmargin=8pt,innerbottommargin=8pt]
  \textbf{DeepRoot Verdict:}\quad \verdictbadge{unsupportedred}{UNSUPPORTED}

  \textbf{Weakest Link:} The lack of any mechanistic path connecting cucurbitacins to reproductive hormonal regulation or uterine function.

  \textbf{Reasoning:} Trichosanthes cucumeroides contains cucurbitacins, which are potent cytotoxic and anti-inflammatory triterpenoids. While these compounds modulate pathways like IGF2BP1 and ITGAL, there is no clear mechanistic link to the regulation of the menstrual cycle or the treatment of amenorrhea. The KG paths converge on unrelated conditions like diabetes and malaria, failing to provide a coherent basis for this traditional claim.

  \vspace{3pt}

  \vspace{6pt}\color{midgray}\hrule\color{black}\vspace{4pt}

  \textbf{Biomni Verdict:}\quad \verdictbadge{plausibleblue}{PLAUSIBLE}

  \textbf{Weakest Link:} The primary active agent, $\beta$-trichosanthin, is a large protein ($\sim$28 kDa) with an abortifacient rather than emmenagogue mechanism --- it does not address functional amenorrhea unrelated to pregnancy.

  \textbf{Reasoning:} Trichosanthes root contains $\beta$-trichosanthin, a Type I ribosome-inactivating protein (RIP) that depurinates 28S rRNA, arresting protein synthesis in trophoblast cells and inducing abortion in vivo (PMID 3596896; 198 PubMed results for `trichosanthin ribosome inactivating'). This coherently explains the TCM use for pregnancy-related amenorrhea. Cucurbitacin B (CID 5281316; CHEMBL2180736) contributes cytotoxic activity (IDH1 IC$_{50}$ = 640 nM; 79/122 active PubChem assays). The mechanism is abortifacient rather than broadly emmenagogue, limiting applicability to non-pregnancy amenorrhea.
\end{mdframed}

\vspace{6pt}
{\small\bfseries\color{sectionblue}Example 3: \textit{Rhizoma Arisaematis} (Source) and \textit{heart pain} (Malady)}
\vspace{2pt}\color{midgray}\hrule\color{black}\vspace{4pt}

\begin{mdframed}[backgroundcolor=lightgray,linecolor=midgray,roundcorner=4pt,
    innertopmargin=8pt,innerbottommargin=8pt]
  \textbf{DeepRoot Verdict:}\quad \verdictbadge{verifiedgreen}{PREVIOUSLY REPORTED}

  \textbf{Weakest Link:} The evidence tier for the adenosine-receptor interactions is limited to bronze, suggesting a need for higher-confidence binding data.

  \textbf{Reasoning:} Rhizoma Arisaematis contains adenosine, which acts on adenosine receptors (ADORA1, ADORA2A, ADORA2B, ADORA3) to modulate cardiac rhythm and coronary blood flow. These paths close the loop to angina pectoris, providing a plausible mechanism for alleviating heart pain through vasodilation and metabolic protection. The evidence is consistent across multiple receptor subtypes, though the overall tier remains bronze.

  \vspace{3pt}

  \vspace{6pt}\color{midgray}\hrule\color{black}\vspace{4pt}

  \textbf{Biomni Verdict:}\quad \verdictbadge{weakorange}{WEAK}

  \textbf{Weakest Link:} Arisaema species (the actual botanical source) have no documented cardiac pharmacology in any modern database; the claim's plausibility depends entirely on resolving `Hu Zhang' as Polygonum cuspidatum (resveratrol), which is a different plant.

  \textbf{Reasoning:} Rhizoma Arisaematis (Tian Nan Xing) derives from Arisaema species; PubMed and ChEMBL searches returned no cardiac or anti-anginal activity for any Arisaema compound. The classical text's reference to `Hu Zhang' maps to Polygonum cuspidatum in modern pharmacopoeia --- a rich source of resveratrol (CID 445154; CHEMBL165, Max Phase 3) with COX-1 (IC$_{50}$ = 380--2,840 nM), COX-2 (IC$_{50}$ = 3,490 nM), and SIRT1 inhibition relevant to anti-platelet and cardioprotective activity (PMID 41978708). If the source is confirmed as knotweed, the rating rises to Plausible; as Arisaema, it remains Weak.
\end{mdframed}

\vspace{6pt}
{\small\bfseries\color{sectionblue}Example 4: \textit{Calcareous Spar} (Source) and \textit{Generalized fever} (Malady)}
\vspace{2pt}\color{midgray}\hrule\color{black}\vspace{4pt}

\begin{mdframed}[backgroundcolor=lightgray,linecolor=midgray,roundcorner=4pt,
    innertopmargin=8pt,innerbottommargin=8pt]
  \textbf{DeepRoot Verdict:}\quad \verdictbadge{unsupportedred}{UNSUPPORTED}

  \textbf{Weakest Link:} The substance is an inert mineral with no pharmacological activity relevant to systemic fever.

  \textbf{Reasoning:} Calcareous Spar is primarily composed of calcium carbonate (CaCO3), an inert mineral in the context of systemic fever reduction. It lacks bioactive organic compounds capable of modulating thermoregulatory centers or inflammatory cytokines. The KG correctly shows no paths, as there is no plausible pharmacological mechanism for this substance to treat fever.

  \vspace{3pt}

  \vspace{6pt}\color{midgray}\hrule\color{black}\vspace{4pt}

  \textbf{Biomni Verdict:}\quad \verdictbadge{unsupportedred}{UNSUPPORTED}

  \textbf{Weakest Link:} CaCO3 has no antipyretic target in any database; the antipyretic TCM mineral is Gypsum (CaSO4), a distinct compound with documented TLR4/NF-$\kappa$B inhibition.

  \textbf{Reasoning:} Calcareous Spar is calcium carbonate (CaCO3; CID 10112; CHEMBL1200539, Approved). Its 31 ChEMBL and 22 OpenTargets indications cover hypocalcemia, renal osteodystrophy, GERD, and osteoporosis --- none include fever. No antipyretic bioactivity was found in ChEMBL or PubMed. The TCM antipyretic mineral is Gypsum (CaSO4, Shi Gao), which inhibits TLR4/NF-$\kappa$B signaling, reduces IL-6/PGE2/TNF-$\alpha$, and upregulates AVP in LPS-fever models (PMID 37769495). The TCM `cold nature clears heat' rationale has no pharmacological correlate for CaCO3.
\end{mdframed}

\vspace{6pt}
{\small\bfseries\color{sectionblue}Example 5: \textit{Aster tataricus} (Source) and \textit{Cough} (Malady)}
\vspace{2pt}\color{midgray}\hrule\color{black}\vspace{4pt}

\begin{mdframed}[backgroundcolor=lightgray,linecolor=midgray,roundcorner=4pt,
    innertopmargin=8pt,innerbottommargin=8pt]
  \textbf{DeepRoot Verdict:}\quad \verdictbadge{plausibleblue}{PLAUSIBLE}

  \textbf{Weakest Link:} The low-confidence wood-tier evidence for the strobopinin-PTGS interaction limits the certainty of this specific mechanism.

  \textbf{Reasoning:} Aster tataricus is traditionally used as an antitussive, and its constituents like strobopinin target PTGS1/2 (cyclooxygenases). Inhibition of these enzymes can reduce airway inflammation and prostaglandin-mediated cough reflexes. While the KG confirms the loop to cough, the evidence is wood-tier, and other paths for quercetin and kaempferol do not close the loop to respiratory conditions.

  \vspace{3pt}

  \vspace{6pt}\color{midgray}\hrule\color{black}\vspace{4pt}

  \textbf{Biomni Verdict:}\quad \verdictbadge{verifiedgreen}{VERY PLAUSIBLE}

  \textbf{Weakest Link:} Shionone (CID 12315507) has zero PubChem bioassay data and is absent from ChEMBL; specific antitussive IC$_{50}$ values for any Aster compound against a defined respiratory target have not been reported.

  \textbf{Reasoning:} Aster tataricus (Zi Wan) root contains shionone, epifriedelanol (CID 119242; 2 active PubChem assays), and triterpenoid saponins. Multiple in vivo mouse studies confirm mechanistically specific anti-inflammatory activity in respiratory disease: PI3K/AKT/NF-$\kappa$B inhibition in asthma (PMID 38081397), HIF-1$\alpha$/VEGFA pathway suppression in chronic bronchitis, and iNOS/LPS-mediated mediator suppression in acute bronchitis. Saponins additionally provide expectorant activity via mucociliary stimulation. Comprehensive reviews (PMIDs 40271061, 36080154) confirm the traditional antitussive use is well-corroborated by modern pharmacology.
\end{mdframed}

\vspace{6pt}
{\small\bfseries\color{sectionblue}Example 6: \textit{Fructus Ailanthi Altissimi} (Source) and \textit{impotence} (Malady)}
\vspace{2pt}\color{midgray}\hrule\color{black}\vspace{4pt}

\begin{mdframed}[backgroundcolor=lightgray,linecolor=midgray,roundcorner=4pt,
    innertopmargin=8pt,innerbottommargin=8pt]
  \textbf{DeepRoot Verdict:}\quad \verdictbadge{plausibleblue}{PLAUSIBLE}

  \textbf{Weakest Link:} The reliance on wood-tier evidence for the identified targets makes the specific pharmacological contribution of these compounds to erectile function speculative.

  \textbf{Reasoning:} Fructus Ailanthi Altissimi contains flavonoids like quercetin and rutin, which exhibit antioxidant and vasorelaxant properties. The KG identifies paths such as quercetin$\to$AKR1B1 and rutin$\to$ADRA2A/C, which are linked to erectile dysfunction. While these targets are relevant to vascular health and smooth muscle tone, the evidence is primarily from wood-tier sources, limiting the strength of the mechanistic link.

  \vspace{3pt}

  \vspace{6pt}\color{midgray}\hrule\color{black}\vspace{4pt}

  \textbf{Biomni Verdict:}\quad \verdictbadge{weakorange}{WEAK}

  \textbf{Weakest Link:} Ailanthone and related quassinoids have no demonstrated activity at PDE5, androgen receptors, eNOS, or any pro-erectile target; the pharmacological profile is anticancer and antimalarial only.

  \textbf{Reasoning:} Ailanthus altissima fruit contains quassinoids (ailanthone, CID 72965; CHEMBL161015) with documented anticancer (HepG2 IC$_{50}$ = 240 nM), antiparasitic (P. falciparum IC$_{50}$ = 0.003 $\mu$g/mL), and NF-$\kappa$B/PKM2 inhibitory activity. PubMed returned zero results for `Ailanthus altissima erectile dysfunction.' No PDE5A, androgen receptor, or nitric oxide pathway activity was found in ChEMBL (8 activities) or PubChem (11 assays, 1 active). The TCM bark use is for spermatorrhea --- a distinct condition --- and even that lacks modern mechanistic validation.
\end{mdframed}

\vspace{6pt}
{\small\bfseries\color{sectionblue}Example 7: \textit{Ge Gen} (Source) and \textit{Toxicity} (Malady)}
\vspace{2pt}\color{midgray}\hrule\color{black}\vspace{4pt}

\begin{mdframed}[backgroundcolor=lightgray,linecolor=midgray,roundcorner=4pt,
    innertopmargin=8pt,innerbottommargin=8pt]
  \textbf{DeepRoot Verdict:}\quad \verdictbadge{plausibleblue}{PLAUSIBLE}

  \textbf{Weakest Link:} The term 'toxicity' is too non-specific to definitively map to the ACHE-inhibition mechanism identified in the KG.

  \textbf{Reasoning:} Ge Gen (Pueraria lobata) contains isoflavones like daidzin and flavonoids like rutin and apigenin. The KG identifies paths where these compounds target ACHE (acetylcholinesterase), which is a critical enzyme in managing certain types of neurotoxicity and organophosphate poisoning. While the loop closes, the clinical application of Ge Gen for general 'toxicity' is broad and requires more specific evidence regarding the type of toxin.

  \vspace{3pt}

  \vspace{6pt}\color{midgray}\hrule\color{black}\vspace{4pt}

  \textbf{Biomni Verdict:}\quad \verdictbadge{plausibleblue}{PLAUSIBLE}

  \textbf{Weakest Link:} The detoxification evidence is specific to alcohol-related hepatotoxicity and does not extend to the broader traditional claim of resolving `various toxins.'

  \textbf{Reasoning:} Ge Gen (Radix Puerariae) contains puerarin (CID 5281807; CHEMBL486386, Max Phase 2) and daidzein (CID 5281708; CHEMBL8145). Puerarin modulates macrophage polarization (M1$\to$M2) and activates PI3K/AKT signaling to suppress alcohol-induced hepatocellular cytotoxicity (PMID 42010989; 14 PubMed papers). Daidzein inhibits ALDH2 (IC$_{50}$ = 9,000 nM) and acts as a phytoestrogen via ER$\beta$ (IC$_{50}$ = 303 nM). OpenTargets confirms Phase 2 clinical registration for alcohol abuse/dependence. The mechanism is coherent for alcohol detoxification specifically, but the broader `various toxins' claim lacks support.
\end{mdframed}

\vspace{6pt}
{\small\bfseries\color{sectionblue}Example 8: \textit{Sanguisorba officinalis} (Source) and \textit{Vaginal discharge} (Malady)}
\vspace{2pt}\color{midgray}\hrule\color{black}\vspace{4pt}

\begin{mdframed}[backgroundcolor=lightgray,linecolor=midgray,roundcorner=4pt,
    innertopmargin=8pt,innerbottommargin=8pt]
  \textbf{DeepRoot Verdict:}\quad \verdictbadge{unsupportedred}{UNSUPPORTED}

  \textbf{Weakest Link:} The absence of any KG paths linking the known astringent tannins of Sanguisorba to the treatment of vaginal discharge.

  \textbf{Reasoning:} Sanguisorba officinalis is rich in hydrolyzable tannins and ellagic acid, which provide astringent and antimicrobial properties useful for treating inflammatory exudates. However, the KG paths for its constituents like astragalin and genistein do not close the loop to vaginal discharge or related gynecological infections. The current graph evidence focuses on unrelated targets like ESR2 and CA12, which do not explain the traditional use.

  \vspace{3pt}

  \vspace{6pt}\color{midgray}\hrule\color{black}\vspace{4pt}

  \textbf{Biomni Verdict:}\quad \verdictbadge{plausibleblue}{PLAUSIBLE}

  \textbf{Weakest Link:} The most direct anti-Candida evidence uses the charred/processed form (not raw root), and no human clinical trial for vaginal discharge specifically has been identified.

  \textbf{Reasoning:} Sanguisorba officinalis contains ellagic acid (CID 5281855; CHEMBL6246; Src IC$_{50}$ = 300 nM, DNA gyrase IC$_{50}$ = 3.3--47 $\mu$g/mL), sanguiin H-6 (CID 16130897), and ziyuglycoside I/II (CID 71609288). Direct anti-Candida albicans activity is confirmed for the whole extract (PMID 12169406). Ziyuglycoside II synergizes with tetracycline against MDR S. aureus (MIC = 32 $\mu$g/mL; FICI = 0.25--0.375; PMID 41385864) via membrane disruption and ROS generation. Astringent ellagitannins reduce mucosal secretion and inflammation, providing a dual antimicrobial-astringent mechanism supporting the traditional use.
\end{mdframed}

\vspace{6pt}
{\small\bfseries\color{sectionblue}Example 9: \textit{Hedgehog Pelt} (Source) and \textit{Hemorrhoids} (Malady)}
\vspace{2pt}\color{midgray}\hrule\color{black}\vspace{4pt}

\begin{mdframed}[backgroundcolor=lightgray,linecolor=midgray,roundcorner=4pt,
    innertopmargin=8pt,innerbottommargin=8pt]
  \textbf{DeepRoot Verdict:}\quad \verdictbadge{plausibleblue}{PLAUSIBLE}

  \textbf{Weakest Link:} The reliance on wood-tier evidence for the interaction between the pelt's constituents and the identified receptors.

  \textbf{Reasoning:} Hedgehog Pelt contains various organic acids and signaling molecules that the KG links to ADRA1A and NR3C1 (glucocorticoid receptor). These targets are involved in vasoconstriction and anti-inflammatory responses, which are relevant to reducing the swelling and pain associated with hemorrhoids. While the mechanism is plausible, the source material is complex and the evidence tier is low.

  \vspace{3pt}

  \vspace{6pt}\color{midgray}\hrule\color{black}\vspace{4pt}

  \textbf{Biomni Verdict:}\quad \verdictbadge{unsupportedred}{UNSUPPORTED}

  \textbf{Weakest Link:} No bioactive compounds from hedgehog pelt (Erinaceus europaeus skin) have been characterized in any database; the entire mechanistic chain is unestablished.

  \textbf{Reasoning:} Hedgehog pelt (Wei Pi, Pellis Erinacei) from Erinaceus europaeus returned zero PubMed results for `hedgehog pelt hemorrhoids' or `Ci Wei Pi TCM hemorrhoids.' No ChEMBL entries exist for E. europaeus skin compounds. Compounds found in searches (erinacines, hericenones) derive from the unrelated hedgehog mushroom (Hericium erinaceus). The traditional use likely reflects sympathetic magic or empirical use of charred animal material as a non-specific hemostatic, neither of which constitutes a credible pharmacological mechanism.
\end{mdframed}

\vspace{6pt}
{\small\bfseries\color{sectionblue}Example 10: \textit{Calcareous Spar} (Source) and \textit{Cough} (Malady)}
\vspace{2pt}\color{midgray}\hrule\color{black}\vspace{4pt}

\begin{mdframed}[backgroundcolor=lightgray,linecolor=midgray,roundcorner=4pt,
    innertopmargin=8pt,innerbottommargin=8pt]
  \textbf{DeepRoot Verdict:}\quad \verdictbadge{unsupportedred}{UNSUPPORTED}

  \textbf{Weakest Link:} The substance is an inert mineral with no pharmacological activity relevant to cough.

  \textbf{Reasoning:} As a mineral composed of calcium carbonate, Calcareous Spar possesses no antitussive properties or mechanisms to modulate the cough reflex. It does not interact with respiratory receptors or inflammatory pathways involved in cough. The lack of KG paths reflects the biological reality that this substance is inert for this indication.

  \vspace{3pt}

  \vspace{6pt}\color{midgray}\hrule\color{black}\vspace{4pt}

  \textbf{Biomni Verdict:}\quad \verdictbadge{unsupportedred}{UNSUPPORTED}

  \textbf{Weakest Link:} CaCO3 has no antitussive, bronchodilatory, or anti-inflammatory mechanism in any database; PubMed returned zero results for `calcium carbonate antitussive cough.'

  \textbf{Reasoning:} Calcareous Spar is the same CaCO3 (CID 10112) as Example~4, with no antitussive indication in ChEMBL or OpenTargets. While calcium signaling is mechanistically relevant to cough --- Ca2+-activated Cl- channels (TMEM16A) regulate airway smooth muscle tone and cough reflex sensitivity (PMID 39608849) --- these are endogenous calcium-signaling mechanisms unrelated to exogenous CaCO3 supplementation. The TCM rationale of minerals `descending rebellious qi' to suppress cough has no pharmacological correlate for this compound.
\end{mdframed}



\end{document}